\documentclass[final,5p,times,twocolumn]{elsarticle}

\usepackage{blindtext, graphicx, amsmath, algorithm, algpseudocode, pifont, algcompatible, comment, layout, amsthm, amssymb}
\usepackage{enumitem}   
\usepackage{eso-pic}
\usepackage{booktabs}
\usepackage{float}

\usepackage[utf8]{inputenc}
\usepackage[english]{babel}
\usepackage{subcaption}
\usepackage{hyperref} 
\hypersetup{ colorlinks=true, linkcolor=black, filecolor=black, urlcolor=cyan, }

\usepackage{caption}
\journal{ICT Express}

\begin{document}

\begin{frontmatter}

\title{Constraint-Compliant Network Optimization through Large Language Models}
\author{Youngjin Song}
\ead{thd4090@korea.ac.kr}

\author{Wookjin Lee}
\ead{mekdugi@korea.ac.kr}

\author{Hong Ki Kim}
\ead{istackcheese@korea.ac.kr}

\author{Sang Hyun Lee\corref{cor1}}
\ead{sanghyunlee@korea.ac.kr}
\address{School of Electrical Engineering, Korea University, Seoul, Korea}

\cortext[cor1]{Corresponding author}

\begin{abstract}
This work develops an LLM-based optimization framework ensuring strict constraint satisfaction in network optimization. While LLMs possess contextual reasoning capabilities, existing approaches often fail to enforce constraints, causing infeasible solutions. Unlike conventional methods that address average constraints, the proposed framework integrates a natural language-based input encoding strategy to restrict the solution space and guarantee feasibility. For multi-access edge computing networks, task allocation is optimized while minimizing worst-case latency. Numerical evaluations demonstrate LLMs as a promising tool for constraint-aware network optimization, offering insights into their inference capabilities.

\end{abstract}

\begin{keyword}
Large language model \sep constraint-compliant linguistic optimization \sep network management
\end{keyword}

\end{frontmatter}


\section{Introduction}\label{sec1}

Future networks are poised to extend the connectivity while fostering the emergence of advanced applications, such as the Internet of Things (IoT) \cite{Ngu}, holographic technology \cite{Pan}, and the metaverse \cite{Wang}. The seamless service of such applications depends on sophisticated network administration, which is realized through efficient network optimization \cite{Zhang}. Driven by rapid evolutions in artificial intelligence (AI), deep learning (DL) has gained significant attention for success across various domains \cite{Goodfellow}. Numerous studies have explored the integration of DL into network applications \cite{Lee}. While DL-based approaches offer notable advantages, they often suffer from limited generalizability and remain constrained to task-specific intelligence \cite{Chen}. 

Large language models (LLMs), exemplified by ChatGPT, have emerged as transformative technologies capable of addressing a wide range of applications \cite{Jiang}. Recent studies have demonstrated that LLMs excel at extracting relevant information from linguistic input and leveraging their internal logic to generate context-aware natural language responses \cite{Zhao}. This inference capability offers the potential for optimization solutions without the need for additional training. Research has explored such problem-solving abilities in domains, including linear regression \cite{Yang} and traveling salesperson problem \cite{Liu}. However, these studies primarily focus on unconstrained optimization problems that allow LLMs to navigate the solution space without predefined restrictions.

When approaching constrained optimization problems with LLMs, constraints are often incorporated into the objective function using penalty functions to transform to an unconstrained formulation \cite{Zhou}. While this relaxation-based approach provides some flexibility in satisfying constraints \cite{Boyd}, it may not be suitable for a class of problems subject to strict requirements.

Motivated by this, we propose a framework that utilizes LLMs for the network optimization while ensuring the constraint satisfaction. The network optimization involves a constrained formulation consisting of an objective function and a set of constraints \cite{Liu3}. The problem structure is converted into a linguistic expression that LLMs can interpret. The proposed approach focuses on restricting the solution space based on given constraints and guides LLMs to explore solutions that meet predefined conditions. Constraints are introduced sequentially, each progressively narrowing the feasible solution space. The final solution is refined from previously explored outputs within the constrained space. 

A feasibility study is conducted using a multi-access edge computing (MEC) network \cite{Liyanage} as an example. In this setup, user devices offload computational tasks to edge servers. By utilizing the computational resources of these servers, the processing latency is reduced, and the overall computational capacity is enhanced. Each user transfers a task to an edge server, where the server processes the computation and returns the results to the user. As servers process tasks from multiple users sequentially, the total processing latency for each server is determined by the sum of the individual user processing times. The total network latency is calculated as the maximum processing time among all servers.

The network latency is evaluated by comparing multiple server latencies. These server latencies are interdependent, as changes in user-server allocations simultaneously affect multiple servers. The proposed approach minimizes the latency through the inference capabilities of LLMs within the MEC network. Numerical results validate the feasibility of the proposed LLM-based optimization, which demonstrates its capability to handle highly complicated solution landscapes. Furthermore, the asymptotic refinement of the framework highlights its potential application in practical network deployment.

The remainder of this paper is organized as follows: Section \ref{sec2} describes the LLM-based network optimization framework. Section \ref{sec3} presents the system model and the application of the proposed framework to the MEC network. Section \ref{sec4} provides numerical results along with the discussion, and Section \ref{sec5} concludes the paper.

\section{LLM-based network optimization framework}\label{sec2}
\begin{figure}
    \centering
    \includegraphics[width=0.85\linewidth]{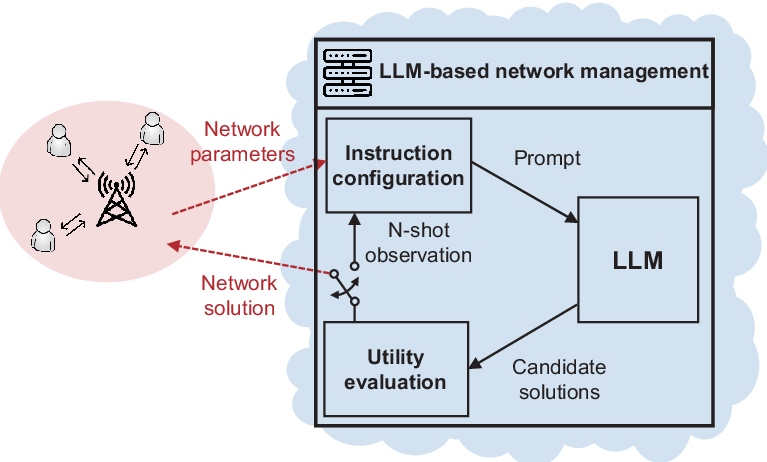}
    \caption{LLM-based network management framework}
    \label{Fig1}
\end{figure}
The overall structure of network optimization is outlined first. A network is characterized by multiple nodes and the connections that represent their interactions. Network optimization focuses on managing these connections to find the best objective value. Network nodes correspond to access points (APs) and their associated users, which are connected via wireless links, while a network controller oversees AP operations. Thus, optimization aims to establish the connection relationship between APs and users. The objective function varies depending on the use case, such as the throughput maximization \cite{Hua} and the latency minimization \cite{Li}, with the corresponding parameters collected by APs. The throughput maximization measures the signal strength between APs and users, whereas the latency minimization focuses on processing delays. The network controller gathers these parameters from APs and processes them to determine the optimal AP assignment for each user.

The network controller equips an LLM agent within a central cloud for network optimization. The LLM agent determines the current network state by obtaining the optimization solution based on network parameters and the previous network state. 
This approach utilizes in-context learning (ICL) \cite{Dong} to refine inferred solutions repeatedly. ICL refers to the capability of a language model to understand contextual information and generate responses based on the provided input. By identifying patterns in previously explored solutions, the LLM agent adapts to various tasks without retraining internal parameters.

Fig. \ref{Fig1} illustrates the proposed LLM-based network optimization framework, which consists of three key components: the instruction configuration, the LLM agent, and the utility evaluation. The instruction configuration component receives the measured network parameters along with an $N$-shot observation, which stores the $N$ most recently inferred solutions and their corresponding objective values. The LLM agent determines the search direction by adjusting its exploration scope based on recurring patterns observed in high-performing solutions. The prompt, which serves as the linguistic input for the LLM agent, is constructed using network parameters and the $N$-shot observation. Based on this prompt, the LLM agent generates new solutions. The utility evaluation component assesses the generated solutions using predefined performance metrics, e.g., throughput and latency, and updates the $N$-shot observation accordingly. Therefore, the proposed framework progressively enhances the optimization performance.

\begin{figure}
    \centering
    \includegraphics[width=0.83\linewidth]{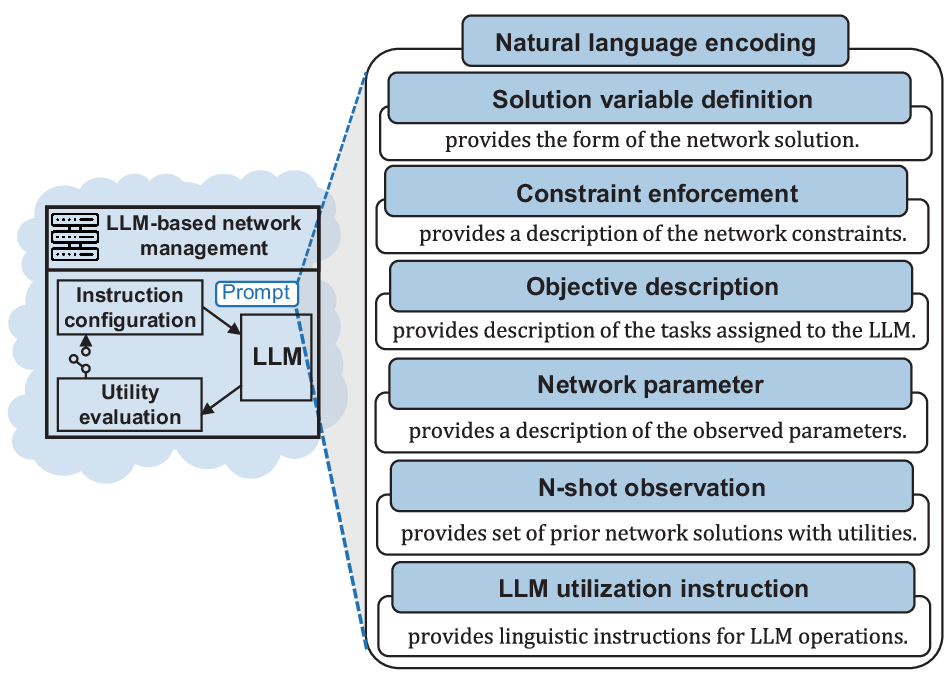}
    \caption{Prompt design for LLM-based network optimization}
    \label{Fig2}
\end{figure}

The prompt is designed to guide the LLM agent in optimizing network connections between APs and users. These connections necessarily comply with network constraints. Each user must be connected to a single AP at a time, and the LLM agent must generate a solution that improves the optimization objective while meeting this requirement. Thus, the prompt structure is explicitly designed to narrow the solution space available to the LLM agent. The solution space consists of all possible AP-user connections, and the network constraints further restrict it to feasible connections. By imposing explicit linguistic constraints, the prompt preemptively filters out infeasible connections so that the LLM agent focuses only on constraint-compliant solutions.

The natural language prompt is structured into six components, as depicted in Fig. \ref{Fig2}: i) The solution variable definition specifies the expected format of the network solution. ii) The constraint enforcement ensures that generated solutions adhere to predefined network regulations. iii) The objective description provides the guidance on the optimization objective. iv) Network parameter input supplies relevant network measurements for the optimization. v) The $N$-shot observation presents previously inferred solutions along with their corresponding objective evaluations. vi) The LLM utilization instruction guides the solution inference, prevents unnecessary code generation, and promotes efficient solution exploration. The prompt structure begins with the solution variable definition and the constraint enforcement, followed by the optimization objective. This organization guarantees that the LLM gains a clear understanding of the feasible solution space before initiating the solution exploration.

To examine the practical feasibility, several implementation considerations are essentially addressed. One key factor is the choice of $N$, which serves as a tunable hyperparameter. A large $N$ allows the LLM agent to obtain rich contextual information, potentially improving the quality of 
inferred solutions. However, increasing $N$ also expands the prompt length, causing high memory consumption and long inference times.
Furthermore, the iterative refinement is carefully regulated to prevent excessive computations. The refinement terminates when one of two conditions is met: i) The solution satisfies a predefined performance criterion, such as the minimum latency requirement or the target throughput level, or ii) the maximum number of iterations is reached. Once either condition is met, the final solution is identified and applied to the network.

\section{Case study: MEC networks}\label{sec3}
This section examines the proposed strategy through a practical application in an MEC network. User devices offload computational tasks to edge servers for processing. Each edge server handles multiple tasks in parallel, and the total network latency is determined by the server with the longest processing time among all servers. 

Minimizing this maximum latency presents a significant challenge since any modification in task distribution simultaneously impacts the workload of multiple servers \cite{Dinh}. Thus, the LLM agent optimizes user-server assignments, which effectively reduces the network latency while satisfying constraints.

\subsection{System model of MEC network}\label{sec3-1}

\begin{figure}
    \centering
    \includegraphics[width=0.9\linewidth]{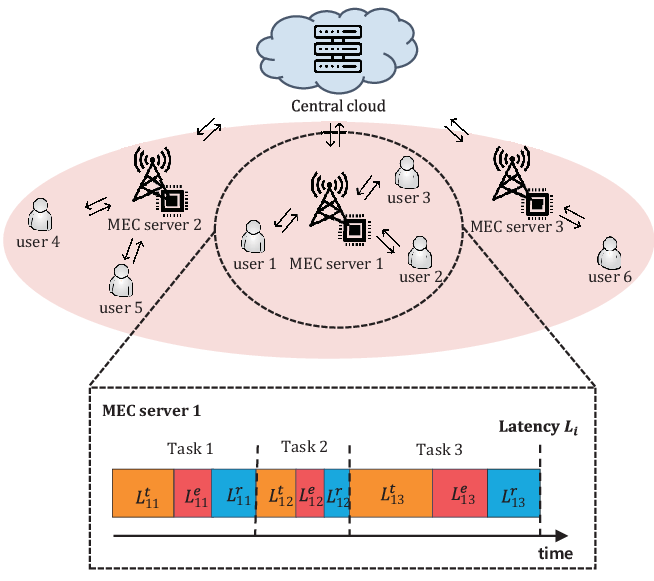}
    \caption{MEC networks and task offloading latency.}
    \label{Fig3}
\end{figure}

Fig. \ref{Fig3} shows a wireless MEC network with $M$ MEC servers, $N$ users, and an LLM agent responsible for managing user-to-server assignments. Each MEC server is equipped with an AP, which establishes connections with users and sequentially processes computational tasks. Let $\mathcal{N} = \{1,...,N\}$ and $\mathcal{M} = \{1,...,M\}$ represent the sets of user tasks and MEC servers, respectively. Task offloading is processed in three phases: i) the task transmission from the user to the MEC server via AP, ii) the task execution on the MEC server, and iii) the retrieval of the processed results by the user through AP.

\begin{figure*}
    \centering
    \includegraphics[width=0.75\linewidth]{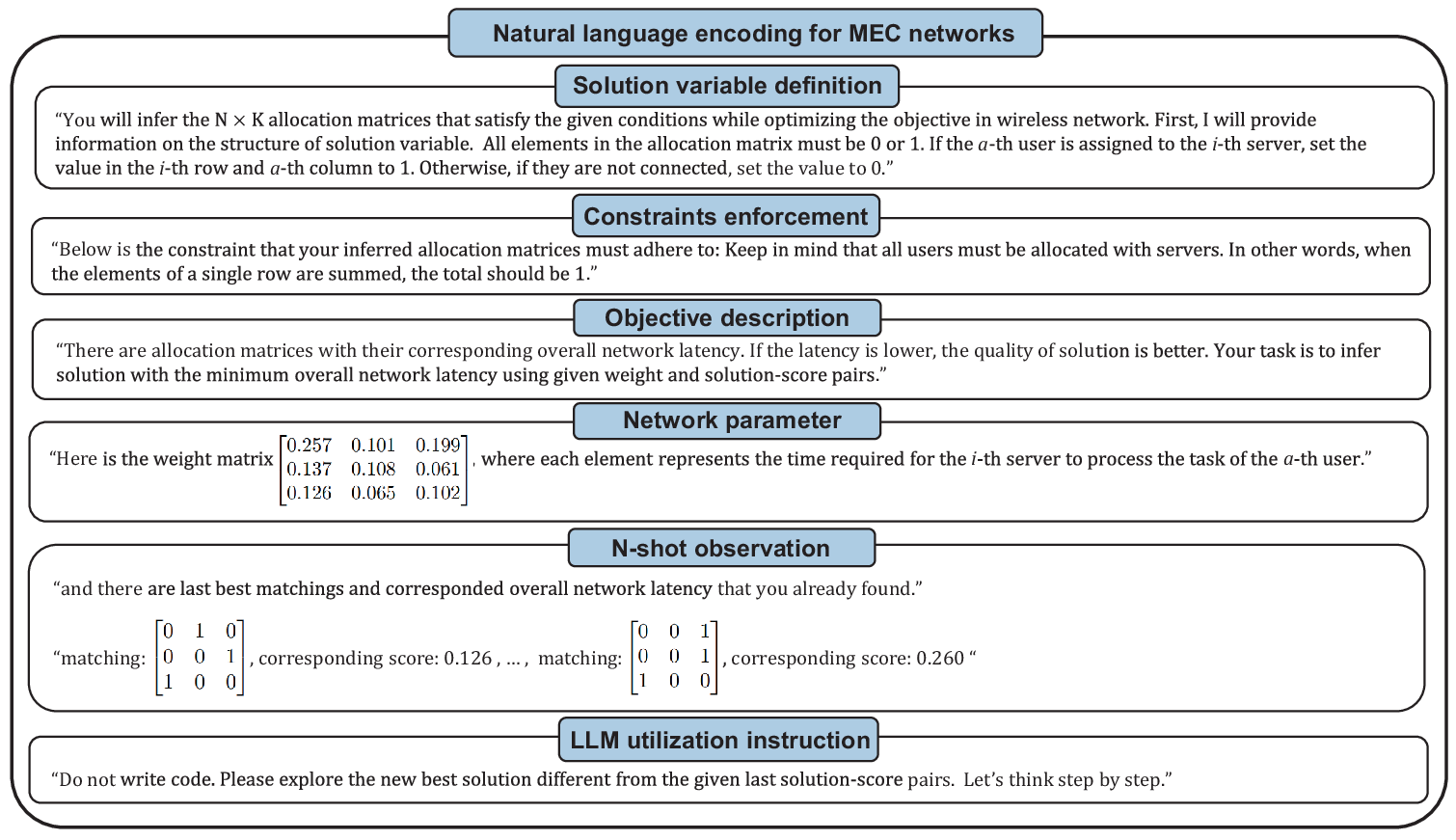}
    \caption{LLM input encoded in natural language for optimization tasks in MEC networks.}
    \label{Fig4}
\end{figure*}

The communication time between the $i$-th MEC server and the $a$-th user during transmission and reception phases depends on the available bandwidth $B_i$, the volume of offloaded and received data, denoted as $D^t_a$ and $D^r_a$, respectively, the transmit power of AP $P_i$ and the user device $P_a$, and the fading channel gain $h_{ia}$. The transmission duration $L^t_{ia}$ represents the time that user $a$ spends to transmit a task to MEC server $i$, while $L^r_{ia}$ is the time for receiving the processed task. These values are given by
\begin{align*}
    L^t_{ia} &= \frac{D^t_a}{B_i\log_2(1 + \frac{P_a h_{ia}}{\sigma^2})},\\
    L^r_{ia} &= \frac{D^r_a}{B_i\log_2(1 + \frac{P_i h_{ia}}{\sigma^2})},
\end{align*}
where $\sigma^2$ is the noise power.
Once the $i$-th AP receives the computation task from $a$-th user, the corresponding MEC server processes it. If the $a$-th task requires $w^t_a$ central processing unit (CPU) cycles to compute $D^t_a$, and the computing capability of the MEC server $i$ is denoted as $f^{\text{mec}}_i$, the computation time elapsed by the task completion of the MEC server $i$ is expressed as
\begin{equation*}
    L^e_{ia} = \frac{w^t_a}{f^{\text{mec}}_i}.
\end{equation*}

The total offloading latency between MEC server $i$ and user $a$ is the sum of $L^t_{ia}$, $L^e_{ia}$, and $L^r_{ia}$, i.e., $L_{ia} = L^t_{ia} + L^e_{ia} + L^r_{ia}$. When multiple users are assigned to a single MEC server, their tasks are processed sequentially. The total processing time $L_i$ required for AP $i$ to handle all assigned tasks is given by 
\begin{align*}
    L_i = \sum_{a \in \mathcal{N}} (L^t_{ia} + L^e_{ia} + L^r_{ia}) x_{ia},
\end{align*}
where $x_{ia}$ is a binary variable indicating whether the task of user $a$ is offloaded to MEC server $i$ ($x_{ia} = 1$) or not ($x_{ia} = 0$). 

Since all edge servers operate in parallel, the total network latency is determined by the server with the longest processing time. Therefore,  the user-to-server allocation problem that minimizes the maximum latency is formulated as 
\begin{subequations}
\begin{align}
    \textbf{P1}: \quad & \min_{x_{ia}} \max_{i} L_i \notag \\
    \text{subject to} \quad & \sum_{i \in \mathcal{M}} x_{ia} = 1, \quad \forall a \in \mathcal{N} \label{eq:constraint1}\\
    & x_{ia} \in \{0, 1\}, \quad \forall (i, a) \in (\mathcal{M}, \mathcal{N}) \label{eq:constraint2}
\end{align}
\end{subequations}
where the constraint in (\ref{eq:constraint1}) dictates that each user is assigned to exactly one MEC server.
The resulting optimization problem \textbf{P1} is formulated in a min-max optimization. Thus, any change in user-server allocations affects multiple servers. Traditional approaches to solving this problem primarily rely on gradient-based algorithms with relaxation techniques \cite{Chen2} or deep learning-based methods \cite{Liu2}. Instead, the proposed approach leverages the inference capability of the LLM agent to solve \textbf{P1}.

\subsection{LLM-based solution for MEC association}\label{sec3-2}
The proposed LLM-based framework addresses the optimization problem \textbf{P1}. For linguistic input processing, each AP measures the task processing time $L_{ia}$ for user $a$ and reports this value to the central cloud, which hosts the LLM agent.
The instruction configuration component composes a prompt using both the $L_{ia}$ values and an $N$-shot observation. Fig. \ref{Fig4} presents a detailed example of the prompt structure tailored for MEC networks. The LLM agent then infers the user-server allocation, which is subsequently evaluated by the utility evaluation component based on latency metrics. This evaluation generates $N$-shot observations that include allocation-latency pairs, which are incorporated into the prompt in subsequent iterations.
This refinement continues until either the allocation inferred by the LLM agent yields a latency below the required threshold for the MEC network operation or the maximum number of iterations is reached. Upon completion, the central cloud transmits the final allocation state to individual APs, and MEC servers allocate computational resources accordingly to users.
The integration of an LLM-based optimization agent within MEC networks enhances the resource allocation efficiency and also provides a scalable approach that can be extended to other network optimization challenges.

\section{Numerical results}\label{sec4}

This section evaluates the proposed framework within the MEC network configurations described earlier. The simulation setup parameters are outlined in Table \uppercase\expandafter{\romannumeral1}. The computational tasks involve applications with predetermined data sizes. The amount of data received by user $a$ after MEC processing is assumed to be $20\%$ of the offloaded data, and the CPU cycles required for processing are given by $w^t_a = 330D^t_a$ \cite{Dinh}. The simulation is conducted within a 1000 m $\times$ 1000 m MEC coverage area, with users uniformly distributed across the region. Gpt-4o-mini is used for its resource efficiency and fast response time, both of which are crucial for real-time MEC network optimization. To prevent uncontrolled growth in the LLM input prompt, the number of $N$-shot examples is fixed at $20$.

\begin{table}[ht]
\centering
\small
\begin{tabular}{l c}
\multicolumn{2}{c}{\textbf{TABLE I}} \\
\multicolumn{2}{c}{\textbf{Simulation Parameters}} \\
\hline
\textbf{Parameters} & \textbf{Value} \\
\hline
Bandwidth of AP ($B_i$) & 10 MHz \\
AP transmit power ($P_i$) & 2 W \\
User device transmit power ($P_a$) & 0.5 W \\
Path loss exponent ($d_e$) & 3\\
Noise power ($\sigma^2$) & $-75$ dBm \\
Required CPU frequency ($w^t_a$) & $330D^t_a$\\
MEC server computing capability ($f^{\text{mec}}_j$) & 1 Gcycles/s \\
Transmitted data size ($D^t_a$) & 5 Mbits \\
Received data size ($D^r_a$) & 1 Mbits\\
\hline
\end{tabular}
\end{table}

\begin{figure}
    \centering
    \begin{subfigure}[b]{0.48\linewidth}
        \centering
        \includegraphics[width=\linewidth]{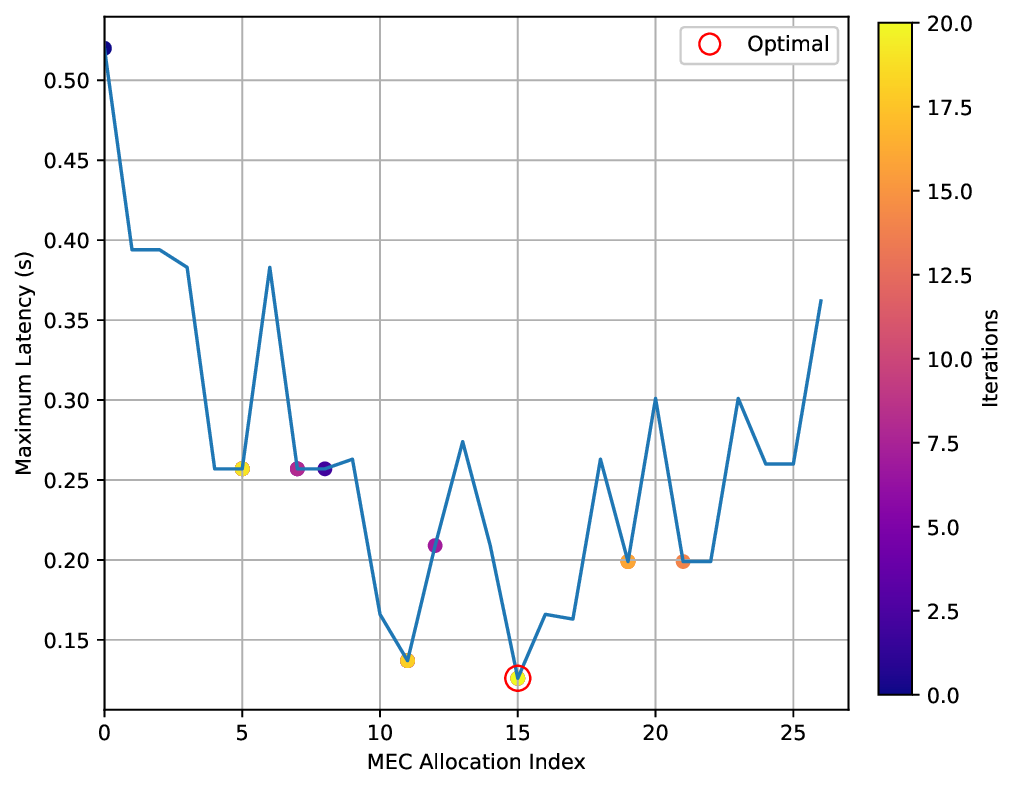}
        \subcaption{ }
        \label{fig5_a}
    \end{subfigure}
    \hfill
    \begin{subfigure}[b]{0.48\linewidth}
        \centering 
        \includegraphics[width=\linewidth]{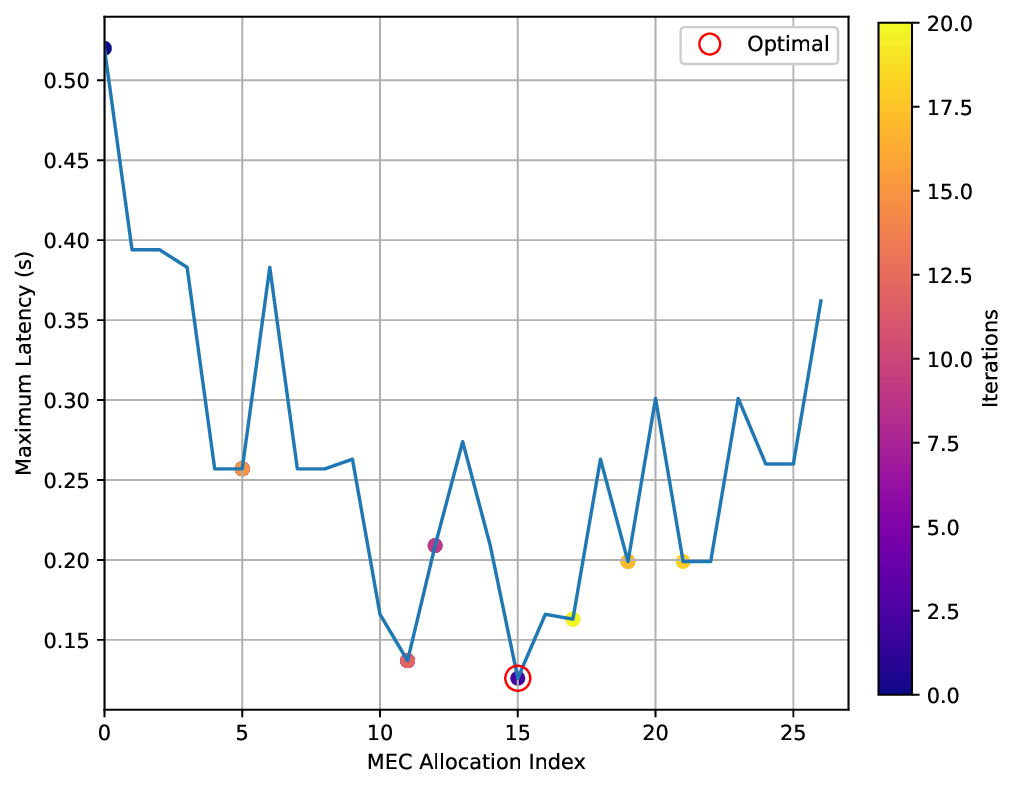}
        \subcaption{ }
        \label{fig5_b}
    \end{subfigure}
    \caption{The solutions identified by the LLM optimizer in the MEC condition where $L_{ia}$ are specified as \eqref{Fig5}.} 
    \label{fig5}
\end{figure}

\begin{figure}
    \centering
    \begin{subfigure}[b]{0.48\linewidth}
        \centering
        \includegraphics[width=\linewidth]{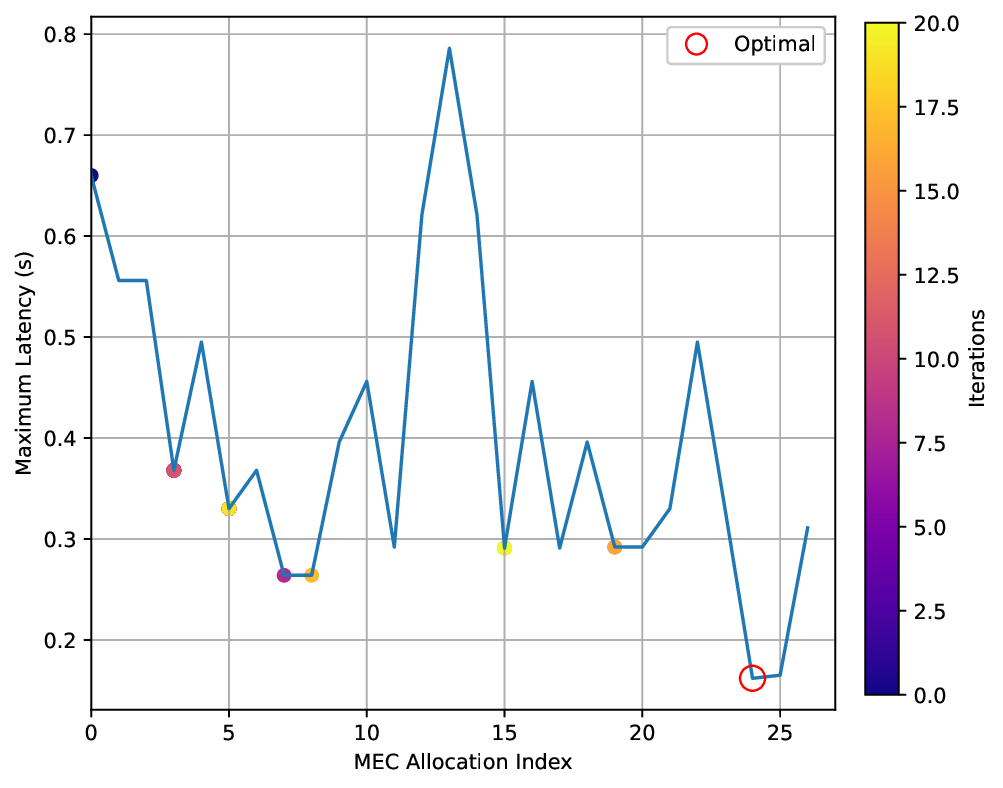}
        \subcaption{ }
        \label{fig6_a}
    \end{subfigure}
    \hfill
    \begin{subfigure}[b]{0.48\linewidth}
        \centering 
        \includegraphics[width=\linewidth]{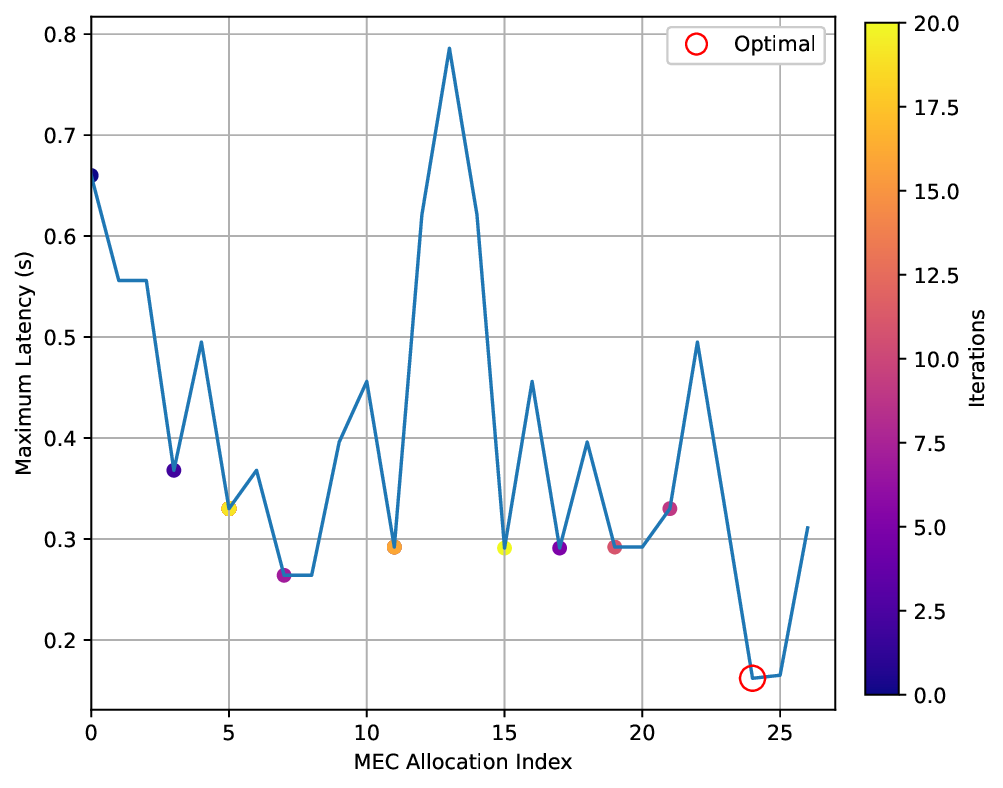}
        \subcaption{ }
        \label{fig6_b}
    \end{subfigure}
    \caption{The solutions identified by the LLM optimizer in the MEC condition where $L_{ia}$ are specified as \eqref{Fig6}.} 
    \label{fig6}
\end{figure}

Fig. \ref{fig5} and Fig. \ref{fig6} illustrate allocation matrices obtained over 20 iterations within the MEC allocation for three MEC servers and three users. The allocation indices are presented comprehensively (see Appendix) with a color bar visually representing the solutions determined by Gpt-4o-mini through the iterations.
For the former case, the link processing times per task $L_{ia}$ are given by
\begin{equation}
   \begin{bmatrix} 
   0.257 & 0.101 & 0.199  \\
   0.137 & 0.108 & 0.061  \\
   0.126 & 0.065 & 0.102  \\
   \end{bmatrix}.
   \label{Fig5}
\end{equation}

For the second case, the link processing times per task $L_{ia}$ are given by
\begin{equation}
   \begin{bmatrix} 
   0.264 & 0.291 & 0.078  \\
   0.292 & 0.330 & 0.084  \\
   0.104 & 0.165 & 0.149  \\
   \end{bmatrix}.
   \label{Fig6}
\end{equation}



The solution space of the MEC problem contains multiple local minima, leading to diverse feasible solutions.
Within this solution space, the proposed LLM-based approach follows distinct search trajectories as the generative model inherently samples outputs in a stochastic manner \cite{Zhao}. Even with probabilistic sampling, the proposed LLM-based approach consistently generates solutions that strictly satisfy the constraint in (\ref{eq:constraint1}). This demonstrates that the proposed prompt structure guides the LLM agent to focus on constraint-compliant solutions. For \eqref{Fig5}, the LLM-based solution consistently identifies the optimal allocation across all instances in Fig. \ref{fig5_a} and Fig. \ref{fig5_b}, despite variations in search trajectories. In contrast, \eqref{Fig6} ends up with a local solution in Fig. \ref{fig6_a} and Fig. \ref{fig6_b}, where one MEC server remains unassigned. Although the proposed method effectively drives the output to a feasible solution that satisfies all constraints, as opposed to other machine learning-based optimization techniques, it may occasionally become trapped at suboptimal solutions out of multiple local minima in case of significant load imbalance across servers. The proposed strategy normally finds the optimal solution in 86.3\% of trials, demonstrating the robustness in navigating complex solution landscapes.

\begin{figure}
    \centering
    \includegraphics[width=0.9\linewidth]{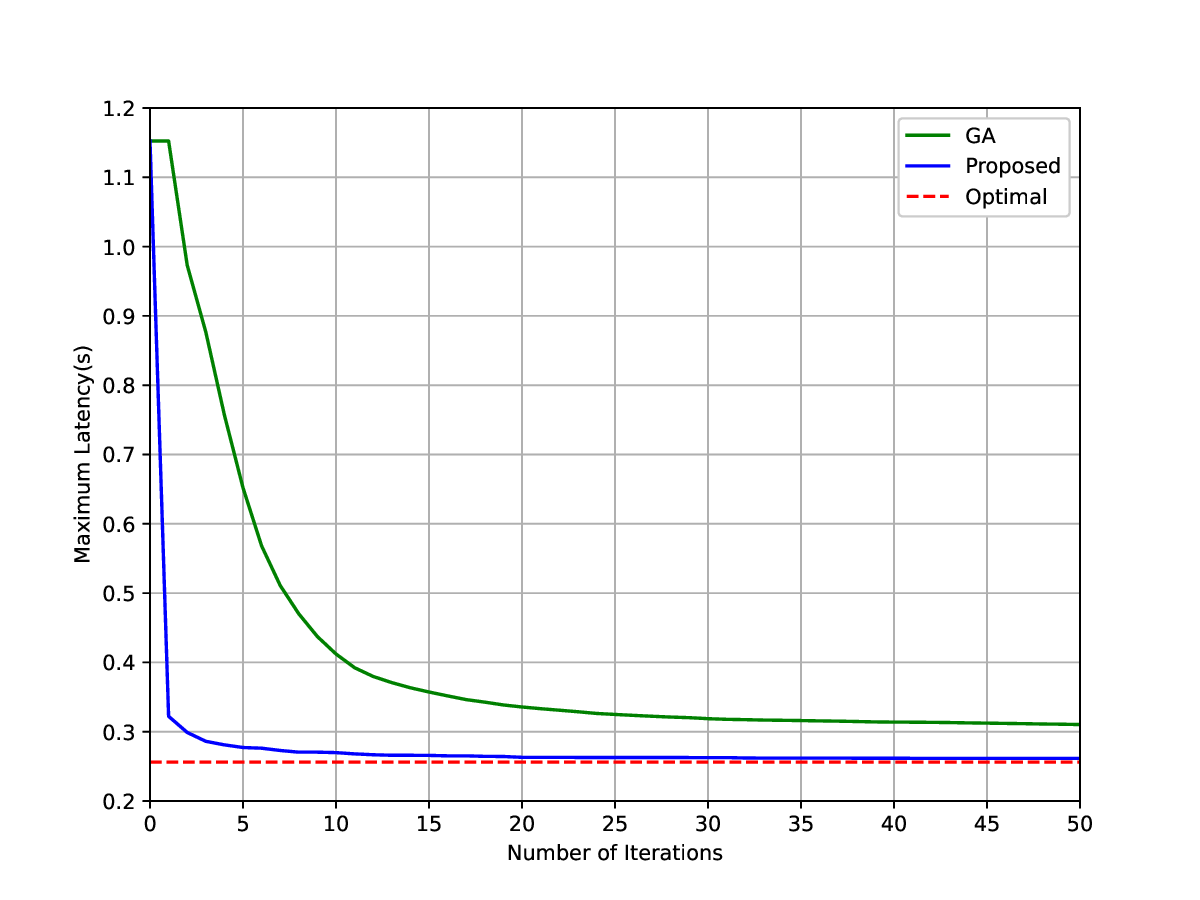}
    \caption{The convergence behavior of the LLM optimizer.}
    \label{iter}
\end{figure}

\begin{figure*}
    \centering
    \begin{subfigure}[b]{0.28\textwidth}
        \centering
        \includegraphics[width=\linewidth]{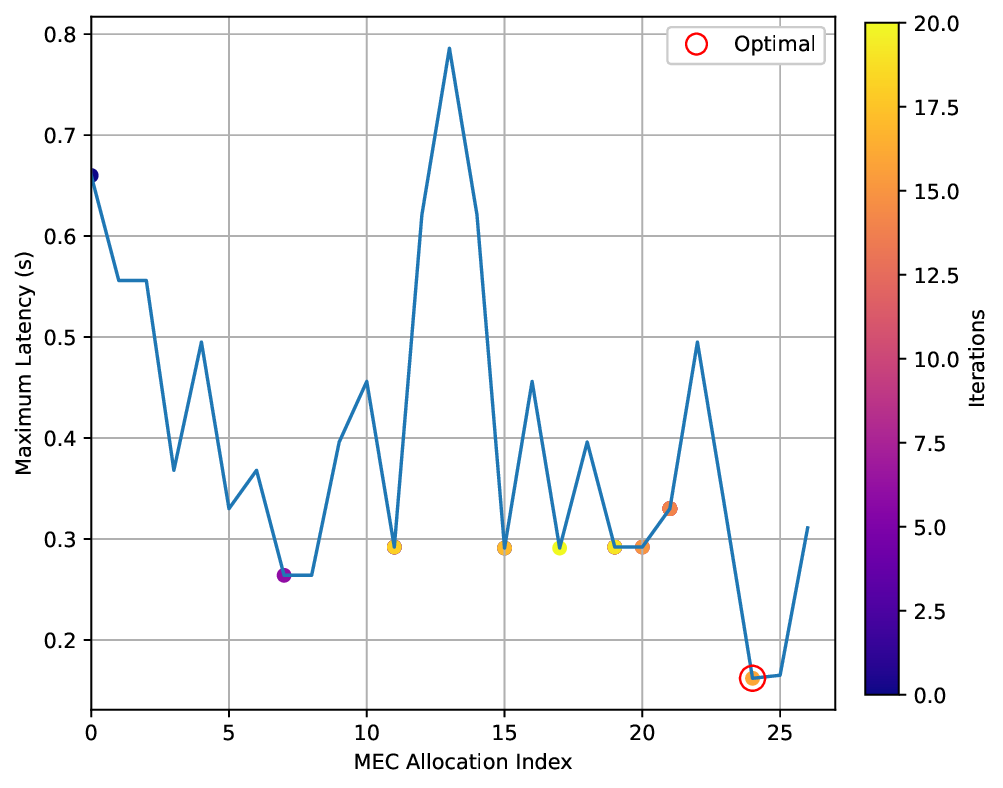}
        \vspace{-20pt}
        \subcaption{}
        \label{fig8_a}
    \end{subfigure}
    \begin{subfigure}[b]{0.28\textwidth}
        \centering 
        \includegraphics[width=\linewidth]{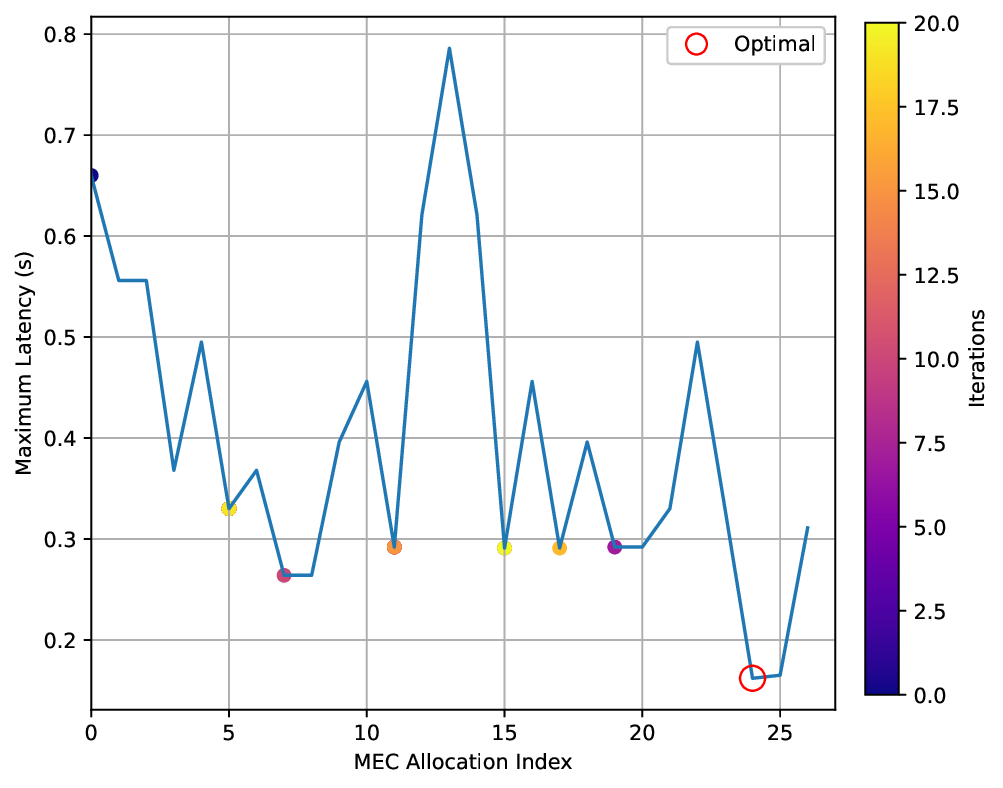}
        \vspace{-20pt}
        \subcaption{}
        \label{fig8_b}
    \end{subfigure}
    \begin{subfigure}[b]{0.28\textwidth}
        \centering 
        \includegraphics[width=\linewidth]{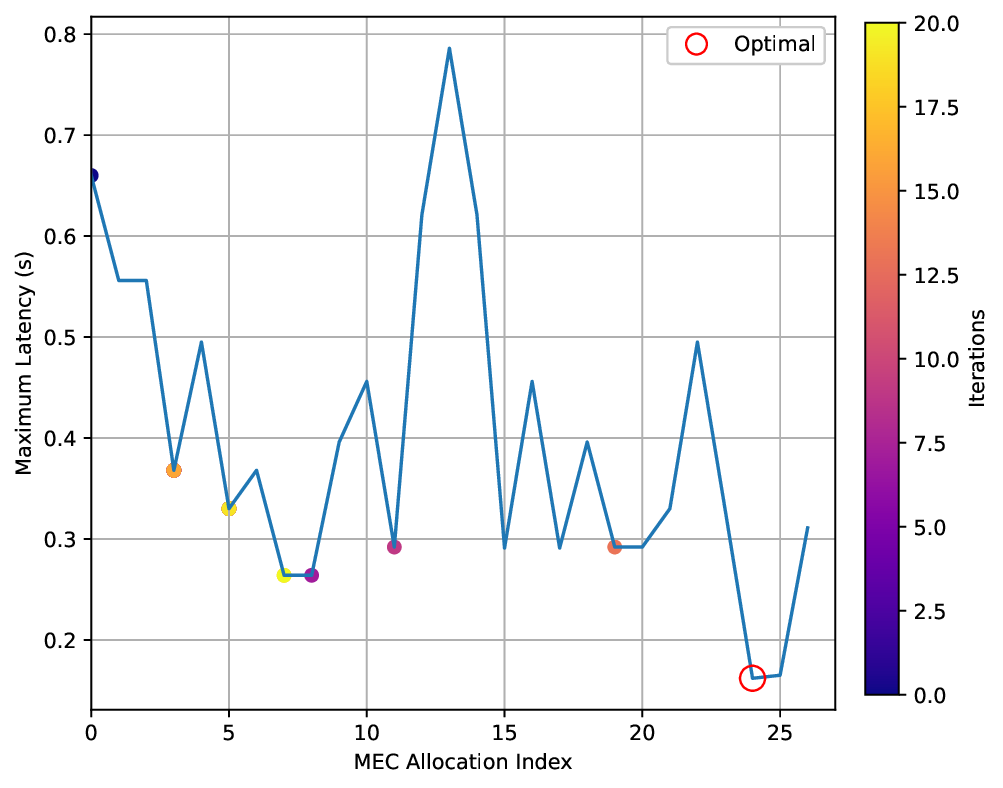}
        \vspace{-20pt}
        \subcaption{}
        \label{fig8_c}
    \end{subfigure}
    \vspace{-3pt}
    \caption{The solutions identified by three LLM optimizers with distinct search trajectories in the MEC condition where $L_{ia}$ are specified as \eqref{Fig6}.}
    \vspace{-3pt}
    \label{fig8}
\end{figure*}


Fig. \ref{iter} illustrates the convergence behavior of the LLM-based framework in an MEC network with three MEC servers and six users.
To assess the performance of the proposed framework, we compare it with the genetic algorithm (GA) \cite{Chien}. 
The GA, a meta-heuristic optimization technique, iteratively searches for solutions by evaluating a population of candidates and refining the most promising allocations. Both the GA and the LLM-based algorithm evaluate five candidate solutions per iteration, leading to a total of 250 possible allocations over 50 iterations. The results reveal that the LLM-based framework outperforms the GA in minimizing maximum latency. In particular, the GA reduces the maximum latency from 1.152s to 0.652s, whereas the LLM-based algorithm reduces it to 0.277s within just five iterations. Furthermore, after 50 iterations, the LLM-based approach converges to an optimal latency within 5 ms, corresponding to a 2\% deviation from the theoretical minimum. This indicates the capability for asymptotic convergence to solutions and the achievement of mostly near-optimal performance.

The numerical results demonstrate that the LLM-based optimization effectively explores complicated solution landscapes and continuously improves its allocations. The current framework relies on a single LLM, which may be occasionally stuck at local minima. To resolve this, multiple LLM agents with diverse search trajectories are deployed to increase the likelihood of reaching the global minimum.
To validate the feasibility, the problem associated with \eqref{Fig6}, where the single LLM agent previously converged to a local minimum, is revisited. Initially 10 different LLM agents are deployed to conduct five preliminary exploration iterations. Then, three agents exhibiting distinct search trajectories are selected for further investigation over the remaining 15 iterations. Figures \ref{fig8_a}, \ref{fig8_b}, and \ref{fig8_c} illustrate the 20-step search trajectories of these selected agents. Fig. \ref{fig8_a} confirms that the resulting search trajectories facilitate to reach the global minimum. Future research is essentially devoted to obtain coordination mechanisms among multiple LLM agents. These mechanisms enable LLM agents to autonomously partition their exploration areas without human intervention and to minimize redundant search trials, while enhancing the diversity in solution exploration.

\section{Conclusion}\label{sec5}

This work develops an LLM-based approach for network optimization. The LLM inference features are used to iteratively refine solutions based on prior evaluations. The linguistic input to the model is carefully crafted to explicitly define the solution structures and enforce necessary constraints, which ensures that the LLM agent operates within a feasible solution landscape. The proposed framework is applied to an MEC network and numerical results demonstrate its practical applicability and effectiveness in optimizing network performance. The proposed approach is shown to deliver efficient solutions while maintaining the flexibility to adapt to dynamic network conditions.

\section*{Acknowledgments}
This work was presented in part, awarded the Best Paper at The 15th International Conference on ICT Convergence (ICTC) and was invited for publication in ICT Express. It was supported in part by the National Research Foundation of Korea (NRF) funded by the Ministry of Science and ICT (MSIT), Korea Government under Grant 2022R1A5A1027646.

\section*{Declaration of competing interest
}
The authors declare that there is no conflict of interest in this paper.

\section*{Appendix}
The appendix provides a detailed explanation of the derivation process for the allocation indices. Let \( n \) be the number of users. We represent the state of each user using a one-hot encoded vector of length 3, equal to the number of MEC servers. Specifically, the vector (1,0,0) indicates allocation to MEC server 1, (0,1,0) represents allocation to MEC server 2, and (0,0,1) denotes allocation to MEC server 3. Given a matrix $\mathbf{X}$ of size \( n \times 3 \) where each row represents the allocation state of a user in one-hot encoding, the objective is to convert this matrix into a single integer. Let \( \mathbf{X} = [x_{ia}] \) be the given matrix where \( i \) ranges from 0 to \( n-1 \) and \( a \) ranges from 0 to 2. Define a mapping function $f: \{(1,0,0), (0,1,0), (0,0,1)\} \to \{0, 1, 2\}$.
The index \( s \) is then computed as
\[
s = \sum_{i=0}^{n-1} f(\mathbf{X}_i) \cdot 3^{n-1-i},
\]
where \( \mathbf{X}_i \) is the \( i \)-th row of matrix \( \mathbf{X} \).

\bibliographystyle{elsarticle-num}

\vspace{-0.3cm}

\begin{thebibliography}{1}

\bibitem{Ngu}
A. H. Ngu, M. Gutierrez, V. Metsis, S. Nepal, and Q. Z. Sheng, ``IoT
middleware: A survey on issues and enabling technologies,'' {\em IEEE
Internet Things J.}, 4(1) (2017) 1–20.

\bibitem{Pan}
Y. Pan, J. Liu, X. Li, and Y.Wang, ``A review of dynamic holographic threedimensional display: Algorithms, devices, and systems,'' {\em IEEE Trans. Ind. Inform.}, 12(4) (2016) 1599–1610.

\bibitem{Wang}
Y. Wang {\em et al.}, ``A survey on metaverse: Fundamentals, security, and privacy,'' {\em IEEE Commun. Surveys Tuts.}, 25(1) (2023) 319–352.

\bibitem{Zhang}
S. Zhang and D. Zhu, ``Towards artificial intelligence enabled 6G: State of the art, challenges, and opportunities,'' {\em Comput. Netw.}, 183 (2020) 107556.

\bibitem{Goodfellow}
I. Goodfellow, Y. Bengio, and A. Courville, {\em Deep Learning.}, MIT Press, (2016).


\bibitem{Lee}
H. Lee, S. H. Lee, and T. Q. S. Quek, ``Deep learning for distributed
optimization: Applications to wireless resource management,'' {\em IEEE J. Sel. Areas Commun.}, 37(10) (2019) 2251–2266.

\bibitem{Chen}
Z. Chen, Z. Zhang and Z. Yang, ``Big AI models for 6G wireless networks: Opportunities, challenges, and research directions,'' {\em IEEE Trans. Wireless Commun.}, 31(5) (2024) 164–172.

\bibitem{Jiang}
F. Jiang {\em et al.}, ``Large language model enhanced multi-agent systems for 6G communications,'' {\em  IEEE Wireless Commun.}, 31(6) (2024) 48–55.

\bibitem{Zhao}
W. X. Zhao {\em et al}., ``A survey of large language models,'' (2023) {\em arXiv:2303.18223}.

\bibitem{Yang}
C. Yang {\em et al.}, ``Large language models as optimizers,''  {\em arXiv:2309.03409}, (2023).

\bibitem{Liu}
S. Liu, C. Chen, X. Qu, K. Tang, and Y.-S. Ong, ``Large language models as evolutionary optimizers,'' in {\em Proc. IEEE Cong. Evol. Comp. (CEC)}, (2024).

\bibitem{Boyd}
S. Boyd, L. Vandenberghe, {\em Convex optimization}, Cambridge University Press, (2004).

\bibitem{Zhou}
H. Zhou {\em et al.}, ``Large language model (LLM) for telecommunications: A comprehensive survey on principles, key techniques, and opportunities", (2024) {\em arXiv:2405.10825}.

\bibitem{Liu3}
Liu {\em et al.,} ``Optimizing wireless systems using unsupervised and reinforced-unsupervised deep learning,'' {\em IEEE network,} 34(3) (2020) 270-277.

\bibitem{Liyanage}
M. Liyanage, P. Porambage, A. Y. Ding, and A. Kalla, ``Driving forces
for multi-access edge computing (MEC) IoT integration in 5G,'' \textit{ICT Exp.}, 7(2) (2021) 127–137.


\bibitem{Hua}
D. T. Hua, Q. T. Do, N. N. Dao, and S. Cho ``On sum-rate maximization in downlink UAV-aided RSMA systems'', \textit{ICT Exp.}, 10(1) (2024) 15-21.

\bibitem{Li}
J. Li, Y. Shi, and Y. Yang, ``Two-stage optimization of computation offloading for ICN-assisted mobile edge computing in 6G network,'' \textit{ICT Exp.}, 11(1) (2025) 26-33.

\bibitem{Dong}
Q. Dong {\em et al.}, ``A survey on in-context learning,'' {\em arXiv:2301.00234} (2024).

\bibitem{Dinh}
T. Q. Dinh, J. Tang, Q. D. La, and T. Q. S. Quek, ``Offloading in mobile edge computing: Task allocation and computational frequency scaling,'' {\em IEEE Trans. Commun.}, 65(8) (2017) 3571–3584.


\bibitem{Chen2}
C. -L. Chen, C. G. Brinton and V. Aggarwal, ``Latency minimization for mobile edge computing networks," {\em IEEE Trans. Mobile Comput,} 22(4), (2023) 2233-2247.

\bibitem{Liu2}
Y. Liu, J. Yan and X. Zhao, ``Deep reinforcement learning based latency minimization for mobile edge computing with virtualization in maritime UAV communication network," {\em IEEE Trans. Veh. Technol.,} 71(4), (2022) 4225-4236.

\bibitem{Chien}
W.-C. Chien, C.-F. Lai, and H.-C. Chao, ``Dynamic resource prediction
and allocation in C-RAN with edge artificial intelligence,'' {\em IEEE Trans. Ind. Informat.}, 15(7) (2019) 4306–4314.

\end{thebibliography}

\end{document}